# The transition temperature of Ising ferromagnetic thin film with uniaxial anisotropy

H. Nakhaee Motlagh[*], H. Moradi

*Department of Physics, School of Sciences, Ferdowsi University of Mashhad, Mashhad, Iran*
(Dated: May 27, 2007)

**Abstract**

The magnetic properties in the Ising ferromagnetic thin films are studied. By transfer matrix method, the transition temperatures are calculated as a function of the intra- and interlayer exchange interactions. The transition temperatures in the Ising ferromagnetic thin films with uniaxial anisotropy are also studied. The results show that the transition temperature changes with the film thickness.

## I. Introduction

Theoretical and experimental studies on the critical theorem in physics, especially in magnetic structures have been of great interest in recent years [1-4]. With the advancement of modern vacuum science and the epitaxial growth technique [5-7], it is now possible to grow very thin films with high quality. The magnetic thin film structures show new fascinating properties which do not exist in the bulk magnetic materials. Investigating the magnetic properties of these films is an important area of research in solid state physics. Studies show that the magnetic properties, such as the magnetization, the transition temperature and *etc*, change with the thickness of the film [8-12]. Experimentally, the thickness-dependent Curie temperature $T_c$ has been measured in Ni [13], Co [14], Fe [15], Gd [16]. In most magnetic thin films, the Curie temperature of the film increases with its thickness. Both theoretical [17-18] and experimental [19-20] investigations reveal that the Ising model is very useful for studying the critical behavior in thin ferromagnetic films.

The main focus of this investigation is to study the Curie temperature based on the mean field theory and transfer matrix method in a thin film that it is assumed as an atomic monolayer. In the present paper, we have completed the theoretical work of Ref. [21].

In section II we present the model and the formalism and will derive the equation that determines the transition temperature. In section III we calculate the transition temperature with uniaxial anisotropy. The results show that the transition temperature $T_C$ is reduced by the finite size effect.

Finally, there is a discussion and a brief conclusion in sections IV and V, respectively.

## II. Model and Formalism

We consider a magnetic thin film as an atomic monolayer with localized spins-1/2. The interaction between the spins is of the ferromagnetic Ising type. The Hamiltonian of the system is given by

$$H = -\sum_{i,j}\sum_{r,r'} J_{i,j} S_{ir} S_{jr'}. \quad (1)$$

Where the summation runs over all pairs of nearest neighbours, $(i, j)$ are plane indices, $(r, r')$ are different sites of the atomic planes, and $S_{ir}$ is spin variable. $J_{ij}$ denotes the exchange constant and is plane dependent.

We obtain spin average as Eq. (2)

$$\langle S_{ir} \rangle = \frac{Tr(S_{ir} \exp(-\beta H))}{Tr(\exp(-\beta H))}. \quad (2)$$

Where $\beta = 1/k_B T$ and $k_B$ being the Boltzman constant. Based on the mean field theory (MFT), in a simplified model, we will solve this problem. However this solution is not exact but usually a qualitative understanding of the critical behavior can be achieved within MFT. For $S=1/2$, Eq. (2) leads to

$$\langle S_{ir} \rangle = \tanh(\beta \sum_{ij} J_{ij} S_{ir}). \quad (3)$$

---

[*] Corresponding author.
E-mail address : ha_nakhaee@yahoo.com( H.Nakhaee)

Near the critical temperature $T_c$, the magnetization is small, and the hyperbolic angel tangent equals with the angle itself, hence Eq. (3) reduces to

$$k_B T m_i = z_0 J_{ii} m_i + z_1 J_{i,i+1} m_{i+1} + z_1 J_{i,i-1} m_{i-1}. \quad (4)$$

Where $m_i$ is the magnetic moment of $i$th thin film layer and $J_{ii}$, $J_{ij}$, $z_0$, $z_1$ and $T$, are the intra-layer exchange coupling, inter-layer exchange coupling, the number of nearest neighbor atoms in plane, the number of nearest neighbor atoms between adjacent planes and the temperature, respectively.

Eq. (4) shows that the magnetization in the $n$th monolayer, $m_n$, depends on the magnetizations in adjacent $(n+1)$th and $(1-n)$th layers. As the temperature reaches higher than the critical temperature $T_C$, the whole system becomes demagnetized and the mean atomic magnetization in every layer approaches zero. Eq. (4) can be rewritten in the following matrical form

$$\begin{pmatrix} m_{i+1} \\ m_i \end{pmatrix} = M_{i-1} \begin{pmatrix} m_i \\ m_{i-1} \end{pmatrix}. \quad (5)$$

With $M_{i-1}$ as the transfer matrix defined by

$$M_{i-1} = \begin{pmatrix} \dfrac{k_B T - z_0 J_{ii}}{z_1 J_{i,i+1}} & -\dfrac{z_1 J_{i,i-1}}{z_1 J_{i,i+1}} \\ 1 & 0 \end{pmatrix} \quad (6)$$

Now let us assume $N_1$ be the number of the atomic monolayers at the right of the origin layer, therefore Eq. (5) can be written as

$$\begin{pmatrix} m_{N_1} \\ m_{N_1-1} \end{pmatrix} = R \begin{pmatrix} m_2 \\ m_1 \end{pmatrix}. \quad (7)$$

Where $R = M_{N_1-1} .... M_2 M_1$ represents successive multiplication of the transfer matrices $M_i$. From Eq. (7) we can write

$$m_{N_1} = R_{11} m_2 + R_{12} m_1 \quad (8)$$
$$m_{N_1-1} = R_{21} m_2 + R_{22} m_1$$

From Eq. (4) we have

$$m_1 \frac{(k_B T - z_0 J_{11})}{z J_{12}} = m_2. \quad (9)$$

Substituting Eq. (9) into Eq. (8) we get

$$R_{11}[(k_B T - z_0 J_{11})/(z J_{11})]^2 \\ + (R_{12} - R_{21})[(k_B T - z_0 J_{11})/(z J_{12})] - R_{22} = 0. \quad (10)$$

The above equation is the general equation for the transition temperature of symmetric films. It is suitable for arbitrary exchange interaction constants $J_{ij}$. Defining $j = z_0 J_{ii}/z_1 J_{ij}$ and $t_C = k_B T/z_1 J_{ij}$ Eq. (10) reduces to

$$R_{11}(t_C - j)^2 + (R_{12} - R_{21})(t_C - j) - R_{22} = 0. \quad (11)$$

$R$ matrix in this state is

$$R = \begin{pmatrix} t - \dfrac{z}{z_0} & -1 \\ 1 & 0 \end{pmatrix}^{N_1-1} \quad (12)$$

Defining $D$ matrix as

$$D = \begin{pmatrix} t - \dfrac{z}{z_0} & -1 \\ 1 & 0 \end{pmatrix} \quad (13)$$

Transfer matrix will be $R = D^{N_1-1}$. Note that $det(D) = 1$, thus $R$ matrix can be written in a linear form as the following

$$R = U_{N_1-1} \begin{pmatrix} t-z/z_0 & -1 \\ 1 & 0 \end{pmatrix} - U_{N_1-2} \begin{pmatrix} 1 & 0 \\ 0 & 1 \end{pmatrix} \quad (14)$$

$$\Rightarrow R = U_{N_1-1} D - U_{N_1-2} I$$

Where $I$ is the identity matrix, $U_N = (\lambda_+^N - \lambda_-^N)/(\lambda_+ - \lambda_-)$, and $\lambda_\pm = (t - \dfrac{z_0}{z} \pm \sqrt{(t-\dfrac{z_0}{z})^2 - 4})/2$ which $\lambda_\pm$ are eigenvalues of $D$ matrix. Substituting Eq. (14) into Eq. (11), we achieve

$$U_{N_1+2} = 0 \quad (15)$$

Choosing $x = (t - z/z_0)/2$ we will have

$$\lambda_+ = x + \sqrt{x^2 - 1} \quad (16)$$
$$\lambda_- = x - \sqrt{x^2 - 1}$$

Hence Eq. (15) becomes

$$U_{N_1+2} = \frac{(x+\sqrt{x^2-1})^{N_1+2} - (x-\sqrt{x^2-1})^{N_1+2}}{2\sqrt{x^2-1}}. \quad (17)$$

If $x^2 > 1$ then a proper choice is $x = \cosh\varphi$. Then Eq. (17) can be writen as

$$U_{N_1+2} = \frac{\sinh[(N_1+2)\varphi]}{\sinh\varphi} \quad (18)$$

And if $x^2 < 1$ with choosing $x = \cos\varphi$, we will arrive at

$$U_{N+1} = \frac{\sin[(N_1+2)\varphi]}{\sin\varphi} \quad (19)$$

From Eq. (18) and Eq. (15) we can obtain the relationship between the magnetic moments at the right of the origin layer as

$$\frac{m_{i+1}}{m_i} = \cos(\frac{\pi}{N_1 + 2}) \qquad (20)$$

If $N_2$ be the number of atomic layers at the left of the origin layer, we get the relationship between the magnetic moments at the left of the origin layer as

$$\frac{m_{i-1}}{m_i} = \cos(\frac{\pi}{N_2 + 2}) \qquad (21)$$

In the above equations we have assumed that the maximum value of the magnetic moment is associated with the central layer. Now we will solve the problem for $N$ magnetic layers which $N = N_1+N_2+1$. Consider that the ratio of the magnetic moments in two successive atomic mono layers in a film with constant thickness are $m_{i+1}/m_i = \alpha_1$ and $m_{i-1}/m_i = \alpha_2$.

$$\frac{k_B T - z_0 J_{ii}}{z_1 J_{i,i+1}} = \alpha_1 + \frac{J_{i,i-1}}{J_{i,i+1}}\alpha_2. \qquad (22)$$

For a film with one atomic mono layer ($N_1=N_2=0$), $\alpha_i=0$ ($i=1,2$). With the assumption $J_{i,i+1}=J_{i,i-1}$, Eq. (22) may be expressed as follows

$$\frac{k_B T - z_0 J_{ii}}{z_1 J_{i,i+1}} = \alpha_1 + \alpha_2. \qquad (23)$$

When the atomic monolayers are symmetric about the central layer ($N_1=N_2=(N-1)/2$), the maximum value of the magnetic moment is corresponded to the central layer of the film. With $m_{i+1}=m_{i-1}$ and $(m_{i+1}/m_i)=(m_{i+1}/m_i)=\alpha_1$, which $m_{i+1}<m_i$, therefore the relationship between the magnetic moments in successive atomic planes can be obtained as

$$m_{i+1} = \cos(\frac{2\pi}{N + 3})m_i. \qquad (24)$$

From Eq. (24) and Eq. (23) we can write the Curie temperature in a thin film as

$$t^c = 2\cos(\frac{2\pi}{N + 3}) + j. \qquad (25)$$

In the limit $N \to \infty$ the bulk reduced Curie temperature is given by

$$t_b^c = 2 + j. \qquad (26)$$

Where $t_b^c = k_B T_{bulk}^c / z_1 J_{i,i+1}$.

### III. The transition temperature of Ising ferromagnetic thin film with uniaxial anisotropy

The Hamiltonian of the Ising ferromagnetic film with aniaxial anisotropy is given by

$$H = -\sum_{i,j} J_{ij} S_i \cdot S_j - \sum_i K_i S_i^2. \qquad (27)$$

Where $K_i$, is the single-ion anisotropy parameter in the atomic monolayer. Using Mean Field Theory spin average is obtained as

$$\langle S_i \rangle = \tanh[\beta(z_0 J_{ii}\langle S_i \rangle + z J_{i,i+1}\langle S_{i+1}\rangle \\ + zJ_{i,i-1}\langle S_{i-1}\rangle + K_i\langle S_i\rangle + K_{i+1}\langle S_{i+1}\rangle \\ + K_{i-1}\langle S_{i-1}\rangle)]. \qquad (28)$$

Near the critical temperature, the order parameter $m_i$ is small, and Eq. (28) reduces to

$$k_B T m_i = (z_0 J_{ii} + K_i)m_i \\ + (z_1 J_{i,i+1} + K_{i+1})m_{i+1} \\ + (z_1 J_{i,i-1} + K_{i-1})m_{i-1}. \qquad (29)$$

In a matrix form the above equation can be written as

$$\begin{pmatrix} m_{i+1} \\ m_i \end{pmatrix} = M_{i-1}\begin{pmatrix} m_i \\ m_{i-1} \end{pmatrix} \qquad (30)$$

With $M_{i-1}$ as the transfer matrix defined by

$$M_{i-1} = \begin{pmatrix} \dfrac{k_B T - (z_0 J_{ii} + K_i)}{z_1 J_{i,i+1} + K_{i+1}} & \dfrac{(z_1 J_{i,i-1} + K_{i-1})}{z_1 J_{i,i+1} + K_{i+1}} \\ 1 & 0 \end{pmatrix} \qquad (31)$$

Non linear equation can be written as

$$R_{11}[\frac{k_B T - (z_0 J_{11} + K_1)}{z_1 J_{12} + K_2}]^2 + (R_{12} - R_{21}) \\ \times [\frac{k_B T - (z_0 J_{11} + K_1)}{z_1 J_{12} + K_2}] - R_{22} = 0. \qquad (32)$$

Therefore the reduced Curie temperature will be

$$t_C = 2\cos(\frac{2\pi}{N + 3}) + \frac{z_0 J_{11} + K}{z_1 J_{12} + K}. \qquad (33)$$

In the limit $N \to \infty$ the bulk reduced Curie temperature is given by

$$t_b^c = 2 + \frac{z_0 J + K}{z_1 J + K}. \qquad (34)$$

When $K = 0$ the above equation reduces to

$$t_C = 2\cos(\frac{2\pi}{N + 3}) + \frac{z_0}{z_1}. \qquad (35)$$

### IV. Results and discussion

From Eq. (24), we can obtain the relationship between magnetic moments of atomic mono layers in a film. Fig (1) shows that $m_{i+1}/m_i$ increases with increasing the number of the atomic monolayers. For $N\to\infty$, $m_{i+1}/m_i \to 1$, it means that the magnetic moment of $m_{i+1}$ is limited to the bulk value of magnetic moment.
The critical temperature $T_C$ is calculated by solving Eq. (25) numerically as a function of the thickness $N$, lattice structure and the coordination numbers ($z_0$, $z_1$). The coordination number is defined as: $z = z_o + 2z_1$, which is different for various structures. For example, for the face-centered cubic with (111) (namely FCC (111)) and (001) (namely FCC(001)) structures, the coordination numbers are ($z_0 = 6$, $z_1 = 3$) and ($z_0 = 4$, $z_1 = 4$), respectively; for a body-centered cubic lattice with (111) structure (namely BCC (111)), $z_0 = 6$, $z = 1$; and for simple cubic lattice, we have $z_0 = 4$, $z = 1$.

In order to illustrate the influence of the lattice structure types, we analyze the dependence of the reduced critical temperature as a function of the thin film thickness $N$. Fig. (2) displays the calculated Cutie temperature $T_C$ as a function of thickness $N$, for FCC (111), BCC (111) and SC(001) structures, note that in all of the structures we assume $J_{11}=J_{12}$.

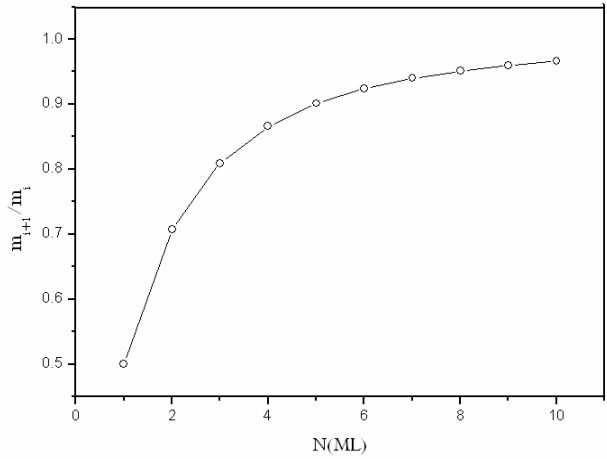

Fig (1). The dependence of $m_{i+1}/m_i$ to $N$.

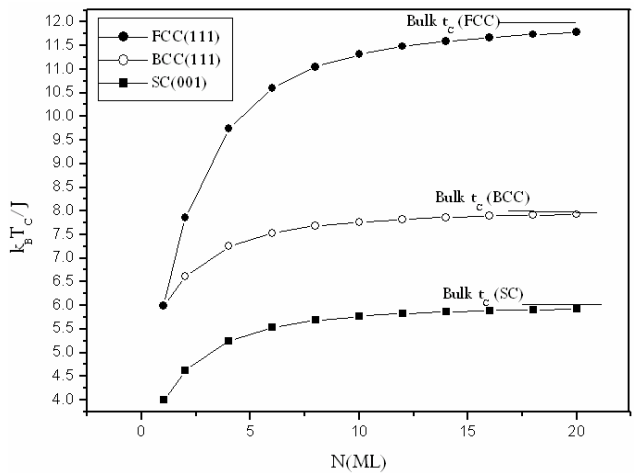

Fig. (2). The Cutie temperatures $T_C$ as a function of thickness $N$, for FCC (111), BCC (111) and SC (001) structures.

In Fig. (3) we consider the variation of $t_C$ with $N$ when $J_{11}/J_{12}$ = 2 and $K/J_{12}$ =1 for four lattice structures namely: FCC (111), FCC (001), BCC (111) and SC (001). It is found that the coordination number of the lattice is responsible for the strong dependence of $T_C$ on the lattice structure. Calculations show that $T_C$ [FCC (111)] >$T_C$ [FCC (001)], because the ratio of the coordination numbers ($z_0$ /$z$) of the FCC (111) ($z_0$ /$z$ =2) structure is larger than the FCC (001) ($z_0$/$z$ =1) structure. This qualitative results for $T_C$ are in agreement with experimental results measured on the thin films of Ni (111) and Ni (001) (FCC lattice structure), i.e., $T_C$ [Ni (111)] >$T_C$ [Ni (001)] [11].

Fig. (4) shows that when K=0 the reduced temperature in FCC (111) structure, is larger than what it is when K≠0. Therefore we can say that the Curie temperature of magnetic thin film decreases when the system has uniaxial anisotropy.

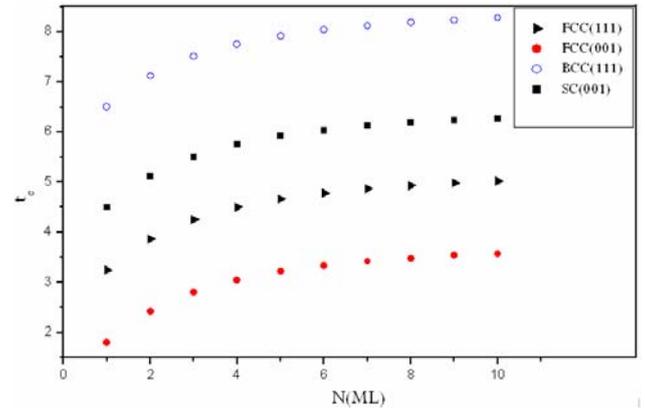

Fig. (3) Thickness dependence of the Curie temperature with ($J_{11}/J_{12}$) = 2 and ($K/J_{12}$) =1 for the four lattice structure namely: FCC (111), FCC (001), BCC (111) and SC (001).

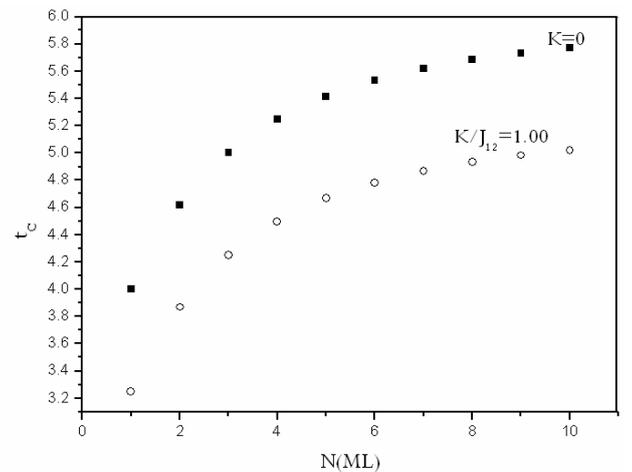

Fig. (5) Thickness dependence of the Curie temperature for FCC (111) lattice, with K=0 and K≠0.

## V. Conclusion

We have studied the transition temperature of Ising magnetic films in SC (001), FCC (111), FCC (001) and BCC (111) structures using the mean field theory and transfer matrix method. Calculations show that the critical temperatures $T_C$ in thin films with various structures changes with increasing the number of layers. The transition temperatures strongly depend on the coordination number in various structures. Using uniaxial anisotropy in the system, the Curie temperature of magnetic thin film will decrease.

### ACKNOWLEDGMENTS

We specially thank Elham Mozaffari for her cooperation.